\documentclass[preprintnumbers,amsmath,amssymb]{revtex4}

\usepackage{graphicx}
\usepackage{dcolumn}
\usepackage{bm}

\begin{document}

\title{Decoherence of two qubits coupled with one-mode cavity without rotating-wave approximation}

\author{Fa-Qiang Wang}
\email{fqwang@scnu.edu.cn}
\author{Wei-Ci Liu}
\author{Rui-Sheng Liang}%
\affiliation{Lab of Photonic Information Technology, School of Information and Photoelectronic Science and Engineering, South China Normal University, Guangzhou 510006, China}%

\date{\today}

\begin{abstract}
The decoherence of two qubits, coupled with one-mode cavity
separately, has been investigated exactly. The results show that,
for the resonant case, the decoherence behavior of system is similar
to Markovian case when the coupling strength is weak, while the
concurrence vanishes in finite time and might recover fractional
initial entanglement before it permanently vanishes
 when the coupling strength is strong. And for detuning case, the entanglement
could periodically recover after a period of time from its
disappearance. These results are quite different from that of system
subjected to Jaynes-Cummings model.
\end{abstract}

\pacs{03.67.Mn,03.65.Yz,03.65.Ud,42.70.Qs}
\maketitle
 \section{Introduction}
\par Recently, many works have been devoted to investigating the
decoherence behavior of entangled qubits, invoked by the phenomenon,
termed as ``entanglement sudden death"(ESD)
\cite{yu2,mf,muh,yu1,mp,ikr,hu,asm,sch,daj,cao,bel,poy,har,wod}. It
is shown that the interaction, between qubits with environment,
could lead to finite-time disentanglement, which is quite different
from the case of continuous variable two-atom model discussed by
Dodd and Halliwell\cite{yu2,dod}.

\par The decoherence of two atoms, coupled with one-mode cavity
separately, has been investigated in Ref.\cite{muh} with
Jaynes-Cummings model, which neglects counter rotating terms
corresponding the emission and absorption of virtual photon without
energy conservation, and is widely used in quantum
optics\cite{Lou,pur,kli}. Generally, the rotating-wave
approximation(RWA), which neglecting counter rotating, is justified
for small detunings and small
 ratio of the atom-field coupling divided by the atomic transition
 frequency. In atom-field  cavity  systems, this ratio is typically of the order  $10^{-7}\sim10^{-6}$.
  Recently, cavity systems with very strong couplings have been discussed\cite{mei}.
 In solid state systems, the ratio may become so large as to consider the effect of counter-rotating wave terms\cite{iri}.
In this paper, we investigate the influence of counter-rotating wave
terms, which are neglected in Ref.\cite{muh}, on the decoherence
behavior of two atoms coupled with one-mode cavity separately.

\par  In section \ref{sec:model}, the reduced non-perturbative quantum master
equation of atom has been derived and its exact solution is
obtained. In section \ref{sec:num}, the decoherence of two initially
entangled qubits has been discussed. The conclusion will be given in
section \ref{sec:clu}.
\section{Model and exact solution}\label{sec:model}
\subsection{non-perturbative master equation and its Exact solution }
\par Now we restrict our attention to two noninteracting two-level
atoms A and B, which are in a perfect one-mode cavity, respectively.
First, we consider the subsystem Hamiltonian of one atom coupled to
one cavity field mode as
\begin{equation}\label{e1}
    H=H_{a}+H_{f}+H_{af}
\end{equation}
where
\begin{eqnarray}
  H_{a} &=& \omega_{0}\frac{\sigma_{z}}{2} \\
  H_{f} &=& \omega a^{\dagger}a \\
  H_{af} &=& g(\sigma_{+}+\sigma_{-})(a^{\dagger}+a)
\end{eqnarray}
where $\omega_{0}$ is the atomic transition frequency between the
ground state $|0\rangle$ and excited state $|1\rangle$.
$\sigma_{z}=|1\rangle\langle1|-|0\rangle\langle0|$,
$\sigma_{+}=|1\rangle\langle0|$ and $\sigma_{-}=|0\rangle\langle1|$
are pseudo-spin operators of atom. $a^{\dagger}$ and $a$ are
creation and annihilation operators of the cavity field mode
corresponding frequency $\omega$. And $g$ is the coupling constant
between the transition $|1\rangle-|0\rangle$ and the field mode.
\par The reduced non-perturbative quantum master
equation of atom could be obtained by path integrals\cite{Ish}

\begin{equation}\label{e5}
\frac{\partial}{\partial t}\rho_{a}= -i \mathcal{L}_{a} \rho_{a} -
\int_{0}^{t} ds \langle \mathcal{L}_{af} e^{-i
\mathcal{L}_0(t-s)}\mathcal{L}_{af}e^{-i
\mathcal{L}_0(s-t)}\rangle_{f}\rho_{a}
\end{equation}

where $\mathcal{L}_0$, $\mathcal{L}_a$ and $\mathcal{L}_{af}$ are
Liouvillian operators defined as

\begin{eqnarray*}
  \mathcal{L}_{0}\rho &\equiv& [H_{a}+H_{f},\rho] \\
  \mathcal{L}_{a}\rho &\equiv& [H_{a},\rho] \\
 \mathcal{L}_{af}\rho &\equiv& [H_{af},\rho]
\end{eqnarray*}
and $\langle...\rangle_{f}$ stands for partial trace of cavity mode.
\par If the cavity field is initially in vacuum state
, the non-perturbative reduced master equation of the atom could be
derived from Eq.(\ref{e5})
\begin{eqnarray}\label{e7}
  \frac{\partial}{\partial
    t}\rho_{a} &=& -g^{2}\left(\alpha^{R}+f(t)\right)\rho_{a} -2i\left(\omega_{0}-g^{2} \alpha^{I}+g^{2}f^{I}(t)\right)J_{0}\rho_{a}\nonumber\\
   & & +g^{2}\left(\alpha+f^{*}(t)\right) J_{+}\rho_{a}+g^{2}\left(\alpha^{*}+f(t)\right)
    J_{-}\rho_{a}+2g^{2}\alpha^{R}K_{+}\rho_{a} \nonumber\\
    & & +2g^{2} \left(\alpha^{R}-f^{R}(t)\right)K_{0}\rho_{a}+ 2g^{2} f^{R}(t)K_{-}\rho_{a}
\end{eqnarray}
Where $J_{0}$, $J_{+}$, $J_{-}$, $K_{0}$, $K_{+}$ and $K_{-}$ are
superoperators defined as
\begin{eqnarray*}
  J_{0}\rho_{a} &\equiv& \left[\frac{\sigma_{z}}{4},\rho_{a}\right] \\
  J_{+}\rho_{a} &\equiv& \sigma_{+}\rho_{a}\sigma_{+} \\
  J_{-}\rho_{a} &\equiv&  \sigma_{-}\rho_{a}\sigma_{-} \\
  K_{0}\rho_{a} &\equiv& (\sigma_{+}\sigma_{-}\rho_{a}+\rho_{a}\sigma_{+}\sigma_{-}-\rho_{a})/2 \\
  K_{+}\rho_{a} &\equiv& \sigma_{+}\rho_{a}\sigma_{-} \\
  K_{-}\rho_{a} &\equiv& \sigma_{-}\rho_{a}\sigma_{+}
\end{eqnarray*}
and
\begin{eqnarray}
  \alpha &=& \frac{1-exp(-i\Delta t)}{i\Delta} \\
  f(t) &=& \frac{exp(i\delta t)-1}{i\delta}
\end{eqnarray}
where $\Delta=\omega+\omega_{0}$, $\delta=\omega_{0}-\omega$.
$\alpha^{R}$, $\alpha^{I}$, $\alpha^{*}$ and $f^{R}$, $f^{I}$,
$f^{*}$ are real part, image part and conjugate of $\alpha$ and of
$f(t)$, respectively.

\par Using algebraic approach, the formal solution of Eq.(\ref{e7}) is
obtained \cite{pur,zhang}
\begin{eqnarray}\label{e8}
  \rho_{a}(t) &=& exp(-\Gamma_{k})\hat{T}exp\left[\int_{0}^{t}dt(\varepsilon_{0}J_{0}+ \varepsilon_{+}J_{+}+\varepsilon_{-}J_{-})\right]\nonumber\\
    & & \times \hat{T}exp\left[\int_{0}^{t}dt(\nu_{0}K_{0}+\nu_{+}K_{+}+\nu_{-}K_{-})\rho_{a}(0)\right]
\end{eqnarray}
where $\hat{T}$ is time ordering operator,
$\varepsilon_{0}=-2i\left(\omega_{0}-g^{2}
\alpha^{I}+g^{2}f^{I}\right) $,
$\varepsilon_{+}=g^{2}\left(\alpha+f^{*}\right)$,
$\varepsilon_{-}=g^{2}\left(\alpha^{*}+f\right)$, $\nu_{0}=2g^{2}
\left(\alpha^{R}-f^{R}\right)$, $ \nu_{+}=2g^{2}\alpha^{R}$,
$\nu_{-}=2g^{2} f^{R}$,
$\Gamma_{k}=g^{2}\left(\tilde{\alpha}^{R}+F^{R}\right)$ and
\begin{eqnarray}\label{e9}
  \tilde{\alpha} &=& \int_{0}^{t}\alpha dt =\frac{1-exp(-i\Delta t)-i\Delta t}{\Delta^{2}} \equiv\tilde{\alpha}^{R}+i\tilde{\alpha}^{I} \nonumber\\
  F &=& \int_{0}^{t}f(t) dt =\frac{1+i\delta t-exp(i\delta t)}{\delta^{2}} \equiv F^{R}+iF^{I}
\end{eqnarray}
where $\tilde{\alpha}^{R}$, $\tilde{\alpha}^{I}$,
$\tilde{\alpha}^{*}$ and $F^{R}$, $F^{I}$, $F^{*}$ are real part,
image part and conjugate of $\tilde{\alpha}$ and of $F$,
respectively.

\par Using the decomposition of SU(2) operator, the exact solution of master
equation Eq.(\ref{e7}) is obtained
\begin{equation}\label{e19}
     \rho_{a}(t) = e^{-\Gamma_{k}}\tilde{\rho}(t)
\end{equation}
\begin{equation}\label{e20}
    \tilde{\rho}(t)=\left(\begin{array}{cc}
                           l\rho^{11}_{a}(0)+ m\rho^{00}_{a}(0) & x\rho^{10}_{a}(0)+ y\rho^{01}_{a}(0) \\
                           q\rho^{01}_{a}(0)+ r\rho^{10}_{a}(0) &
                           n\rho^{00}_{a}(0)+ p\rho^{11}_{a}(0)
                         \end{array}
                         \right)
\end{equation}
\begin{eqnarray}
  l &=& e^{k_{0}/2}+e^{-k_{0}/2}k_{+}k_{-},\ m = e^{-k_{0}/2}k_{+} \\
  n &=& e^{-k_{0}/2},\ p= e^{-k_{0}/2}k_{-}\\
  q &=& e^{-j_{0}/2},\ r= e^{-j_{0}/2}j_{-} \\
  x &=& e^{j_{0}/2}+e^{-j_{0}/2}j_{+}j_{-},\ y=e^{-j_{0}/2}j_{+}
\end{eqnarray}
where $j_{+}$, $j_{0}$, $j_{-}$ and $k_{+}$, $k_{0}$, $k_{-}$
satisfy the following equation\cite{pur}
\begin{eqnarray}
  \dot{X}_{+} &=& \mu_{+}-\mu_{-}X_{+}^{2}+\mu_{0}X_{+} \\
  \dot{X}_{0} &=& \mu_{0}-2\mu_{-}X_{+}\\
  \dot{X}_{-} &=& \mu_{-}exp(X_{0})
\end{eqnarray}
$\mu=\varepsilon$ for $X=j$ and $\mu=\nu$ for $X=k$.

\subsection{ Concurrence}
\par Throughout the paper, we use Wootters concurrence\cite{woott}. For simplicity, we
assume that the two subsystems have the same parameters. The
concurrence of the whole system could  be obtained\cite{ikr}
\begin{eqnarray}
  C_{\xi} &=& max \left\{0,c_{1},c_{2}\right\},(\xi=\Phi,\Psi) \\
  c_{1} &=& 2e^{-2\Gamma_{k}}(\sqrt{\rho_{23}\rho_{32}}-\sqrt{\rho_{11}\rho_{44}}) \nonumber\\
  c_{2} &=&
  2e^{-2\Gamma_{k}}(\sqrt{\rho_{14}\rho_{41}}-\sqrt{\rho_{22}\rho_{33}})\nonumber
\end{eqnarray}
corresponding to the initial states of $|\Phi\rangle =
\beta|01\rangle+\eta|10\rangle$ and $|\Psi\rangle =
\beta|00\rangle+\eta|11\rangle$, respectively. Where $\beta$ is real
and $0<\beta<1$, $\eta=|\eta|e^{i\varphi}$ and
$\beta^{2}+|\eta|^{2}=1$. The reduced joint density matrix of the
two atoms, in the standard product basis
$\mathcal{B}=\{|1\rangle\equiv|11\rangle, |2\rangle\equiv|10\rangle,
|3\rangle\equiv|01\rangle, |4\rangle\equiv|00\rangle \}$, could be
written as\cite{bel}
\begin{equation}\label{e28}
    \rho^{AB}=e^{-2\Gamma_{k}}
    \left(\begin{array}{cccc}
            \rho_{11} & 0 & 0 & \rho_{14} \\
            0 & \rho_{22} & \rho_{23} & 0 \\
            0 & \rho_{32} & \rho_{33} & 0 \\
            \rho_{41} & 0 & 0 & \rho_{44}
          \end{array}
          \right)
\end{equation}
here the diagonal elements are
\begin{eqnarray*}
  \rho_{11} &=& l^{2}\rho_{11}(0)+lm\rho_{22}(0)+ml\rho_{33}(0)+m^{2}\rho_{44}(0) \\
  \rho_{22} &=& lp\rho_{11}(0)+lm\rho_{22}(0)+mp\rho_{33}(0)+mn\rho_{44}(0) \\
  \rho_{33} &=& lp\rho_{11}(0)+pm\rho_{22}(0)+nl\rho_{33}(0)+nm\rho_{44}(0) \\
  \rho_{44} &=& p^{2}\rho_{11}(0)+pn\rho_{22}(0)+np\rho_{33}(0)+n^{2}\rho_{44}(0)
\end{eqnarray*}
and the nondiagonal elements are
\begin{eqnarray*}
  \rho_{14} &=& x^{2}\rho_{14}(0)+xy\rho_{23}(0)+yx\rho_{32}(0)+y^{2}\rho_{41}(0) \\
  \rho_{23} &=& xr\rho_{14}(0)+xq\rho_{23}(0)+yr\rho_{32}(0)+yq\rho_{41}(0) \\
  \rho_{32} &=& rx\rho_{14}(0)+ry\rho_{23}(0)+qx\rho_{32}(0)+qy\rho_{41}(0) \\
  \rho_{41} &=& r^{2}\rho_{14}(0)+rq\rho_{23}(0)+qr\rho_{32}(0)+q^{2}\rho_{41}(0)
\end{eqnarray*}

 \section{Numerical results and discussion}
 \label{sec:num}

 \par In order to compare the results with that of two Jaynes-Cummings atoms in Ref.\cite{muh},
we primarily investigate the decoherence for resonant case
$\omega=\omega_{0}$.

\par
\begin{figure}
  \includegraphics{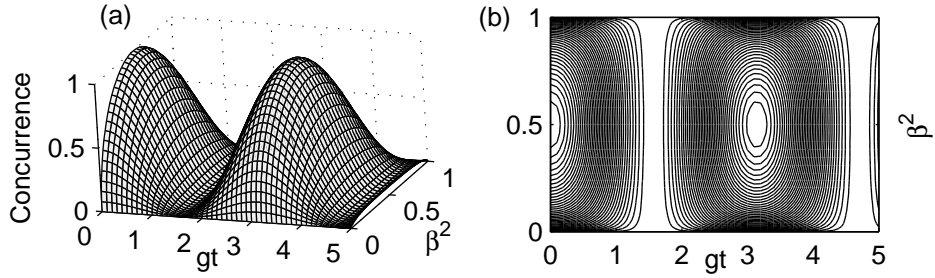}\\
  \caption{$C_{\Phi}$ and its contour for resonanct case as a function of $gt$ and $\beta^{2}$ with RWA.}\label{F1}
\end{figure}

\begin{figure}
  \includegraphics{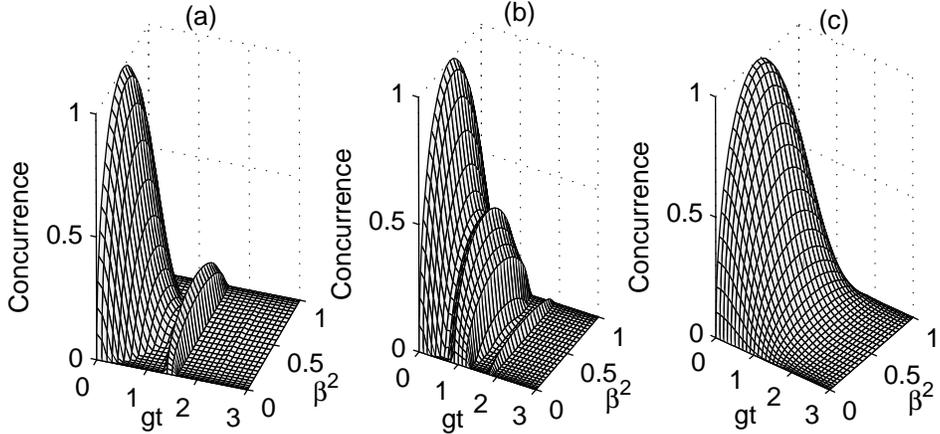}\\
  \caption{$C_{\Phi}$ as a function of $gt$ and $\beta^{2}$ for $\delta=0$. (a) $\omega_{0}=1.5g$, (b)$\omega_{0}=3g$, (c)$\omega_{0}=30g$.}\label{F2}
\end{figure}

 First, we focus on the decoherence of
 two qubits with initial state of $|\Phi\rangle$. For the RWA model
in Ref.\cite{muh}, Fig.\ref{F1} shows that the concurrence
periodically vanishes and revives. And the change of $C_{\Phi}$
against $\beta^{2}$ is symmetrical because of the symmetry of the
initial state $|\Phi\rangle$. The decoherence, of the non-RWA model
in this paper, has been shown in Fig.\ref{F2}.

\par (A) As $\omega_{0}=1.5g$, Fig.\ref{F2}(a) reveals that
 the concurrence $C_{\Phi}$ decreases
to zero in a finite time, vanishes for a period of time, revives
with small amplitude and then vanishes permanently. This
characteristic will hold on for more stronger coupling constant.

\par (B) From Fig.\ref{F2}(b), we
could find that, when $\omega_{0}=3g$, the concurrence $C_{\Phi}$
first decreases to a certain value and maintain it for a period of
time before it vanishes, like a rumple, then it revives with a
smaller amplitude than that in Fig.\ref{F2}(a) and vanishes
permanently. As the coupling constant $g$ decreases, there are more
rumples and a smaller amplitude revival.

\par (C) When $\omega_{0}=30g$, Fig.\ref{F2}(c) exhibits that the concurrence $C_{\Phi}$ decreases monotonically and
 exponentially to zero without revival of entanglement. This
characteristic will hold on for more weaker coupling constant.

\par
\begin{figure}
  \includegraphics{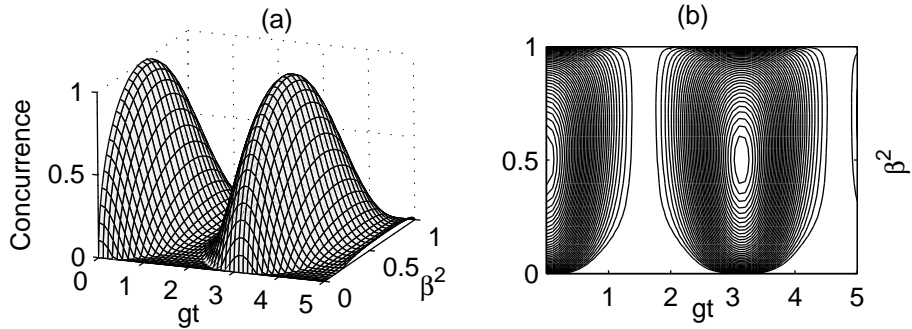}\\
  \caption{$C_{\Psi}$ and its contour as a function of $gt$ and $\beta^{2}$ with RWA.}\label{F3}
\end{figure}

 Then, we focus on the decoherence of
 two qubits with initial state of $|\Psi\rangle$. For the RWA model
in Ref.\cite{bel}, Fig.\ref{F3} shows that the entanglement
represented by $C_{\Psi}$ can fall abruptly to zero, and will remain
zero for a period of time before it recovers. The length of time
interval for the zero entanglement is dependent on the degree of
initial entanglement. The time interval for $\beta^{2}<1/2$ is
longer than that for $\beta^{2}>1/2$. There are periodical
disappearance and revival of entanglement in time scale. For the
non-RWA model, the decoherence of the system was also categorized
into three cases
\par
\begin{figure}
  \includegraphics{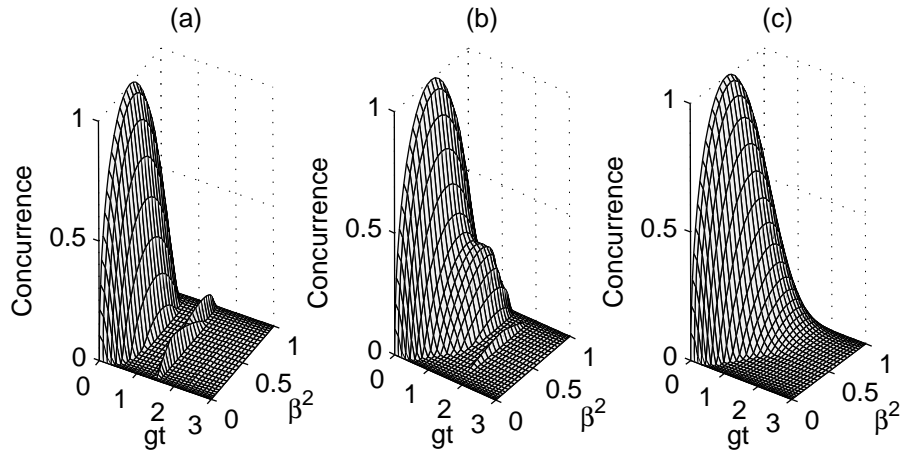}\\
  \caption{$C_{\Psi}$ as a function of $gt$ and $\beta^{2}$ for $\delta=0$. (a) $\omega_{0}=2g$, (b)$\omega_{0}=3.5g$, (c)$\omega_{0}=40g$.}\label{F4}
\end{figure}

 (A) When $\omega_{0}=1.5g$, Fig.\ref{F4}(a) reveals that, for $\beta^{2}>1/2$, the concurrence $C_{\Psi}$ decreases
exponentially to zero, remain zero for a period of time, revives
fractional initial entanglement and then vanishes permanently, while
$C_{\Psi}$ vanishes permanently after a finite time for
$\beta^{2}<1/2$, similar to the Markovian case\cite{yu2}.
\par (B) As $\omega_{0}=3g$, Fig.\ref{F4}(b) shows that the entanglement
represented by $C_{\Psi}$ has a similar behavior to that of
$C_{\Phi}$ for $\beta^{2}>1/2$ in Fig.\ref{F2}(b). In contrast, for
 $\beta^{2}<1/2$, it decreases smoothly to zero at finite time, then vanishes permanently after a small
 amplitude entanglement revival link that for $\beta^{2}>1/2$.

\par (C) When $\omega_{0}=30g$, Fig.\ref{F4}(c) exhibits that it first decreases to zero at short time, then vanishes
permanently after a small amplitude entanglement revival. Unlike the
two cases above, the evolution behavior of concurrence $C_{\Psi}$
becomes symmetric, like that of $C_{\Phi}$, because of the strong
interaction of atom with reservoir through the emission and
absorption of virtual photon.

\par From Fig.\ref{F2} and Fig.\ref{F4}, we find that the entanglement will
 decreases to zero finally for no-RWA and there are no periodical disappearance and revival of entanglement,
like that in Fig.\ref{F1} and Fig.\ref{F3} for RWA case. It is the
enhancement of spontaneous emission, as an atom resonantly coupled
with a cavity, leads to the disappearance of entanglement\cite{kle}.
We could conclude that the neglect of counter-rotating wave terms in
Hamiltonian of RWA model leads to the different characteristics of
$C_{\Phi}$.

\begin{figure}
  \includegraphics{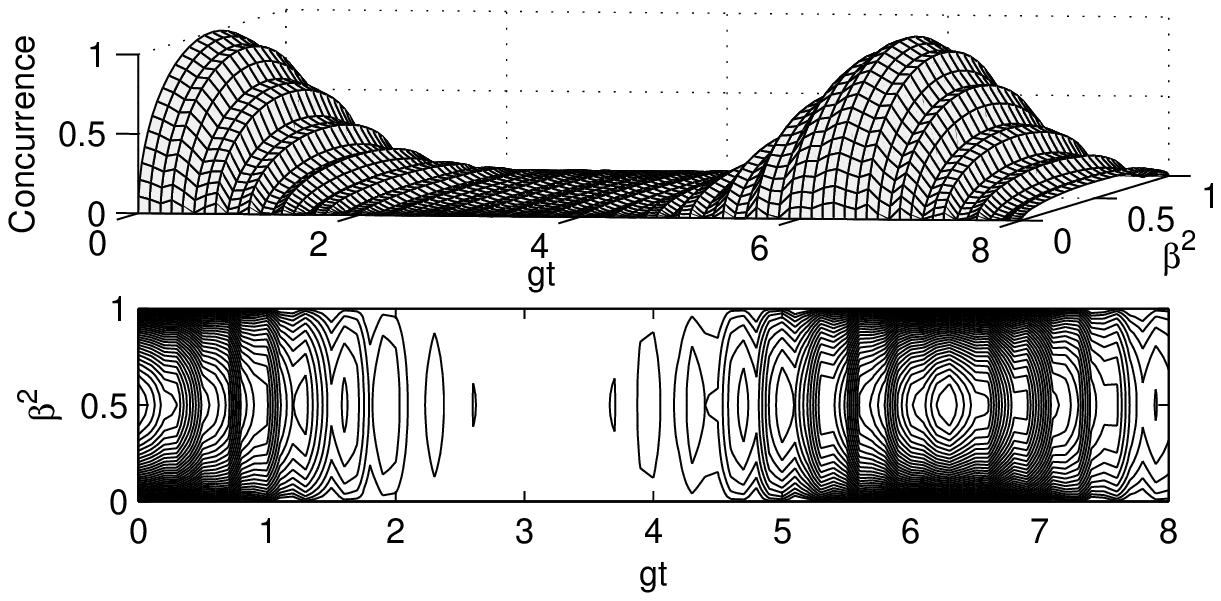}\\
  \caption{Concurrence $C_{\Psi}$ and its contour for non-resonant case as a function of $gt$ and $\beta^{2}$ with $\omega_{0}=10g, \delta=0.1\omega_{0}$.}\label{F5}
\end{figure}

\begin{figure}
  \includegraphics{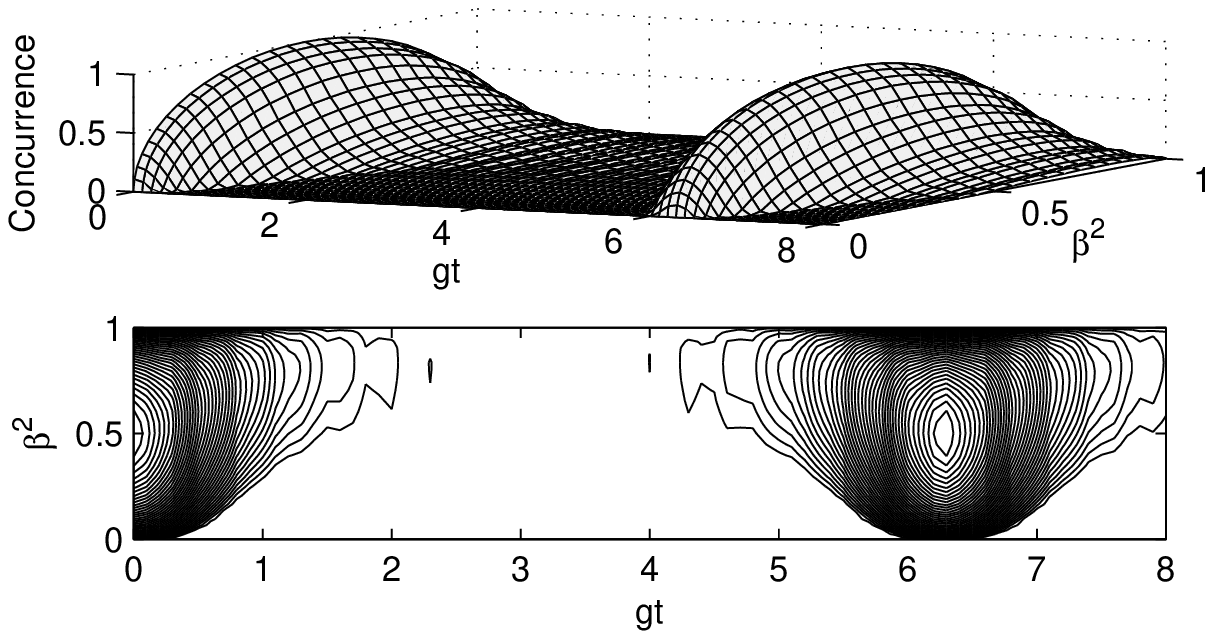}\\
  \caption{Concurrence $C_{\Psi}$ and its contour for non-resonant case as a function of $gt$ and $\beta^{2}$ with $\omega_{0}=10g, \delta=0.1\omega_{0}$.}\label{F6}
\end{figure}

\par Finally, we investigate whether there are periodical disappearance
and revival of entanglement for detuning case. Concurrence
$C_{\Phi}$ and $C_{\Psi}$ for detuning case as a function of $gt$
and $\beta^{2}$ are shown in Fig.\ref{F5} and Fig.\ref{F6},
respectively.

\par From Fig.\ref{F5} and Fig.\ref{F6}, we could find that the
initial entanglement will recover after a period of time from its
disappearance because the spontaneous emission could be greatly
inhibited, as an atom  non-resonantly coupled with cavity
mode\cite{kle}. The revival time interval is dependent on the
detuning $\delta$. The bigger the detuning is, the shorter the time
interval is. We also find that the entanglement will change little
when the detuning is large and coupling strength is weak.

\section{Conclusion}\label{sec:clu}
\par The decoherence of two initially
entangled atoms, coupled with two one-mode cavities separately, has
been discussed exactly. The results show that the decoherence
behavior of two qubits is dependent on the coupling strength and the
detuning between atom transition frequency and the cavity mode.

Firstly, there are no periodical disappearance and revival of
entanglement for resonant case like that for Jaynes-Cummings model
in Ref.\cite{muh}. Secondly, for detuning case, the entanglement
could periodically recover after a period of time from its
disappearance. Thirdly, for resonant case, the decoherence behavior
of system is similar to Markovian case when the coupling strength is
weak, while the concurrence will vanishes in finite time and might
recover fractional initial entanglement before it permanently
vanishes
 when the coupling strength is strong.
\par The results also
exhibit that the RWA in Hamiltonian leads to the existence of
revival of entanglement for resonant case and it might be improper
to take RWA when the interaction between atom and external field is
correlated and strong.

\section*{Acknowledgments} This work was supported by the State
Key Program for Basic Research of China under Grant No.2007CB307001
and by the Science and Technology of Guangdong Province, China(Grant
No.2007B010400066).

\newpage 
\bibliography{apssamp}

\end{document}